# Network Topologies of Financial Market During the Global Financial Crisis

Ashadun Nobi[1,2], Seong Eun Maeng[1], Gyeong Gyun Ha[1], Jae Woo Lee[1]

[1]Department of Physics, Inha University, Incheon 402-751 South Korea

[2]Department of Computer Science and Telecommunication Engineering, Noakhali Science and Technology University, Sonapur Noakhali-3802, Bangladesh

The global financial crisis that exploded when Lehman Brothers declared bankruptcy has been heavily influenced all over the world[1-3]. Before the 2008 global financial crisis, the subprime mortgage crisis had swept the American housing market and the world financial markets in 2007[4,5]. The accumulated pressures building in the subprime market burst partly in 2007 before finally erupting the following year. The impact on global finance spreads simultaneously all over the world[4-7]. The damages in emerging markets were particularly hard, similar to a giant earthquake. We consider the effects of the global financial crisis through a local Korean financial market around the 2008 crisis[4]. We analyze 185 individual stock prices belonging to the KOSPI (Korea Composite Stock Price Index), considering three time periods: the time before, during, and after the crisis. The complex networks generate from the fully connected correlation network by using the cross-correlation coefficients among the stock price time series of the companies. We generate the threshold networks (TN), the minimal spanning tees (MST), and the hierarchical network (HN) from the fully connected cross-correlation networks[8-10]. By assigning a threshold value of the cross-correlation coefficient, we obtain the threshold networks. We observe the power law of the degree distribution in the limited range of the threshold. The degree distribution of the largest cluster in the threshold networks during the crisis is fatter than other periods. The clustering coefficient of the threshold networks follows the power law in the scaling range. We also generate the minimal spanning trees from the fully connected correlation networks[8]. The MST during the crisis period shrinks in comparison to the periods before and after the crisis. The cophenetic correlation coefficient increases during the crisis, indicating that the hierarchical structure increases during this period[11]. When the crisis hit the market, the companies' behave synchronously and their correlations become stronger than the normal period

Different models, techniques, and theoretical approaches have been developed to describe the features of financial dynamics[10-15]. Network techniques have become important tools for describing and quantifying complex systems in many branches of science[16-18]. More recent works have been devoted to topology characteristics, clustering, and community structure in networks[16-18]. The MST has been applied to financial markets such as the Dow Jones Industrial Average (DJIA) and the Standard & Poor's 500[10]. During the last decade, the correlation and the network analysis in many financial markets have been successfully investigated[15-19]. The network topology of the German stock market around the global financial crisis has also been studied recently[20]. At around the same time, the network structures in a financial market changed heavily due to the large fluctuations of the market dynamics[20,21]. We focus on the impact of the crisis caused by different network techniques on the network structure in the Korean stock market. In order to build a natural network, we consider the cross-correlations of the stock price fluctuations. We analyze and compare the topological properties, the degree distribution and the network structure by the threshold method throughout the cycle of the crisis in 2008[22]. The MST technique for stocks, called the asset tree, has been studied in order to reflect the financial market taxonomy around the recent worldwide financial crash[8,7]. A hierarchical method is also used to produce a tree-like dendrogram to reinforce the network properties[19-25].

To understand the effects of the global financial crisis on the network structures, we selected over 185 companies belonging to the KOSPI 200 from June 2, 2006 to December 30, 2010. We divide the full time series into three time windows based on the fluctuation of volatility, specifically the periods before, during, and after the crisis. The probability distribution functions of the cross-correlation coefficients around the 2008 financial crisis deviate from the Gaussian function (Fig. 1). During the crisis the probability distribution becomes broader than in the other periods, indicating a wide range of higher cross-correlations coefficient during the crisis (Table S1). The mean cross-correlation coefficient during the crisis is higher than in other periods. We consider the period after the crisis as a normal due to it have the lowest volatility (Table S1). The deviation of the cross-correlation before the crisis from normal periods provides a red flag for the coming stock market crash. The lower value of the mean cross-correlation coefficient after the crisis indicates that the market is in a more stable state than in any other periods.

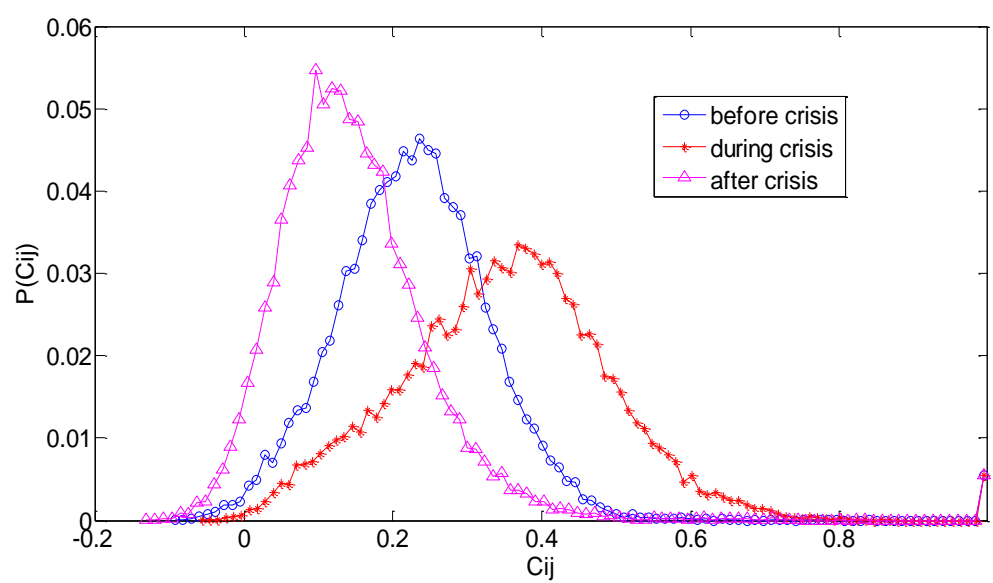

Figure 1 | Probability density functions of the cross-correlation coefficients calculate using the daily returns of the 185 individual stock prices in the Korean stock market before, during, and after the crisis. The distribution functions become broader and the mean increases during the crisis.

We generate a threshold network by assigning a threshold value of the cross-correlation coefficients. In the threshold network, a vertex (V) represents a distinct company and an edge (E) represents a cross-correlation value of the return time series between the companies. We specify a certain threshold θ, $-1 \leq \theta \leq 1$, for the cross-correlation coefficients. If the correlation coefficient $C_{ij}$ is greater than or equal to θ, we add an undirected edge connecting the vertices i and j. So, different values of θ define the networks with the same set of vertices, but different sets of edges[8]. We construct a threshold network with the threshold θ as the average value of the cross-correlation coefficient plus the integer multiplication of the standard deviation (Fig. 2). The numbers represent the companies belong to the KOSPI 200. We gave the full name of the companies in the supplementary Excel file.

The largest clusters in the threshold network show different morphological structures depending on the threshold and the observed period. At a threshold where $\theta = \overline{C_{ij}} + \sigma$ (0.3313, 0.4665 and 0.2508 before, during, and after the crisis, respectively), most of the companies belong to the largest cluster (Fig.2 a, d, g), and the percentages of the vertices in the largest cluster are 83%, 79.5%, 85.4% before, during, and after the crisis, respectively. The absolute value of the threshold during the crisis is higher than in other periods, but the clustering percentage is closer to other periods, which indicates a strong correlation among the indices during the crisis. The companies are less densely connected if we increase the threshold value. At the threshold, $\theta = \overline{C_{IJ}} + 2\sigma$, the clustering percentages are 52%, 42.7% and 49.1% before, during, and after the crisis, respectively. The interesting feature the companies' reorganization can be found in this financial network at threshold $\theta = \overline{C_{IJ}} + 3\sigma$. Here, many small clusters sit besides the largest one. We present the clusters having vertices more than three (Fig. 2c, f, i). We observe that the number of the small clusters during the crisis is the only one having two vertices (not shown in figure). However, the percentage of the clustering of the vertices to the largest one (20%) is higher than those before the crisis (4.8%) and after the crisis (18.9%). Most of the companies appear then to belong to the large group that resisted crisis shock during the crisis. Moreover, we observe the obvious clustering of the companies to the sectors at this threshold. Before the crisis, there are six clusters where the largest one belongs to the financial sector. The rest of the clusters are concentrated on the sector of heavy industry and shipbuilding i.e. construction (red square), finance (green triangle), iron, and metal products (magenta plus) and non-metallic mineral products (blue star) (Fig. 2c, f, i). When the crisis was happening, the clusters tied to finance and heavy industry and shipbuilding combine to form a strongly linked cluster. After the crisis, the largest one consists of the financial and heavy industry and shipbuilding sectors while the second and third biggest comes from electronics, automobile, and transport equipment (diamond orange), respectively. Before the 2008 financial crisis, the failures of the financial systems became noticeable in the financial sector like the subprime mortgage market. The problems in the market create strong correlations among companies in the financial sector (Fig. 2c). When the crisis happened, there are strong connections between heavy industry and engineering (Fig. 2f and Fig. 2i). In these periods, the finances and the heavy industry and engineering constitute the strong community structures.

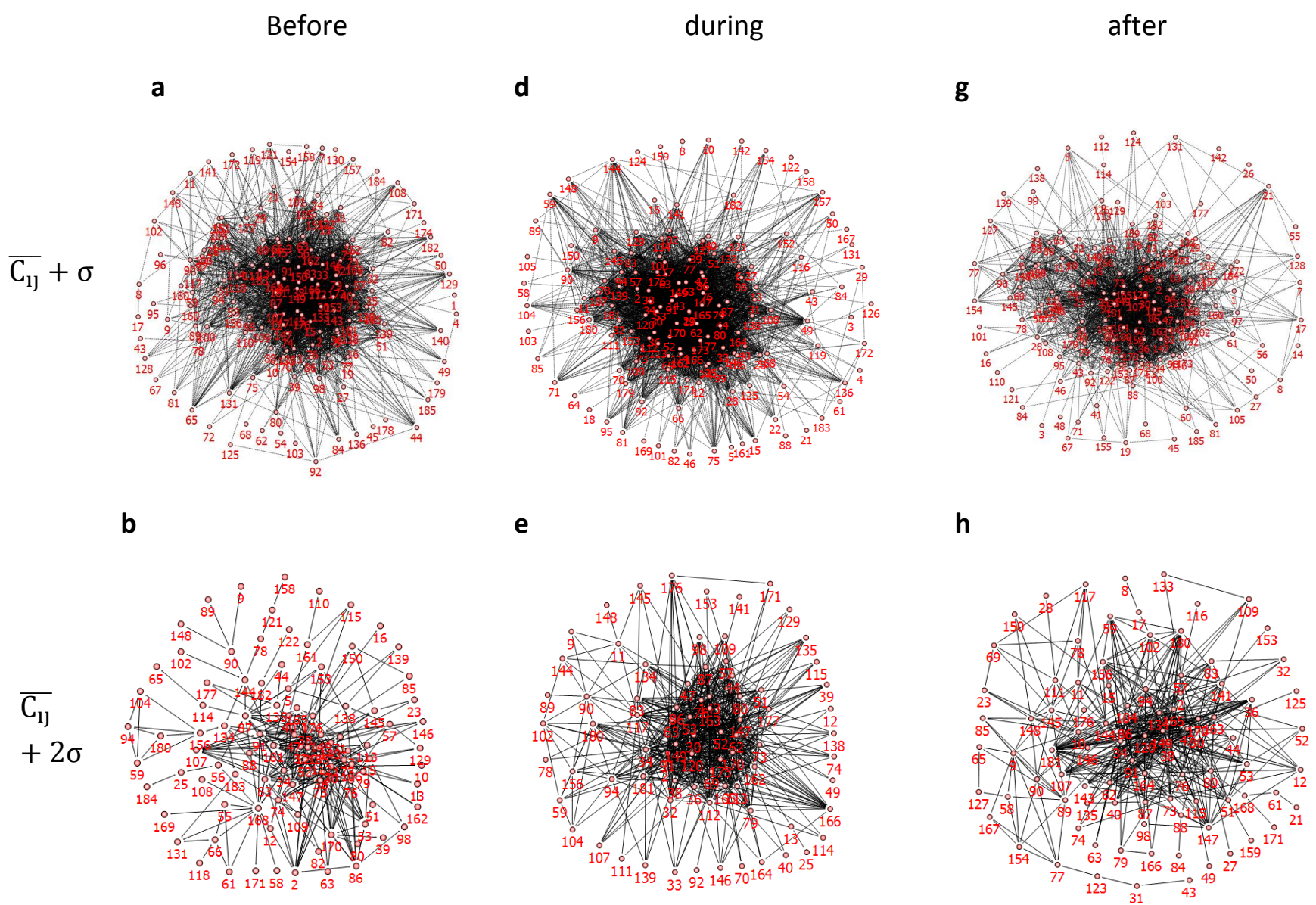

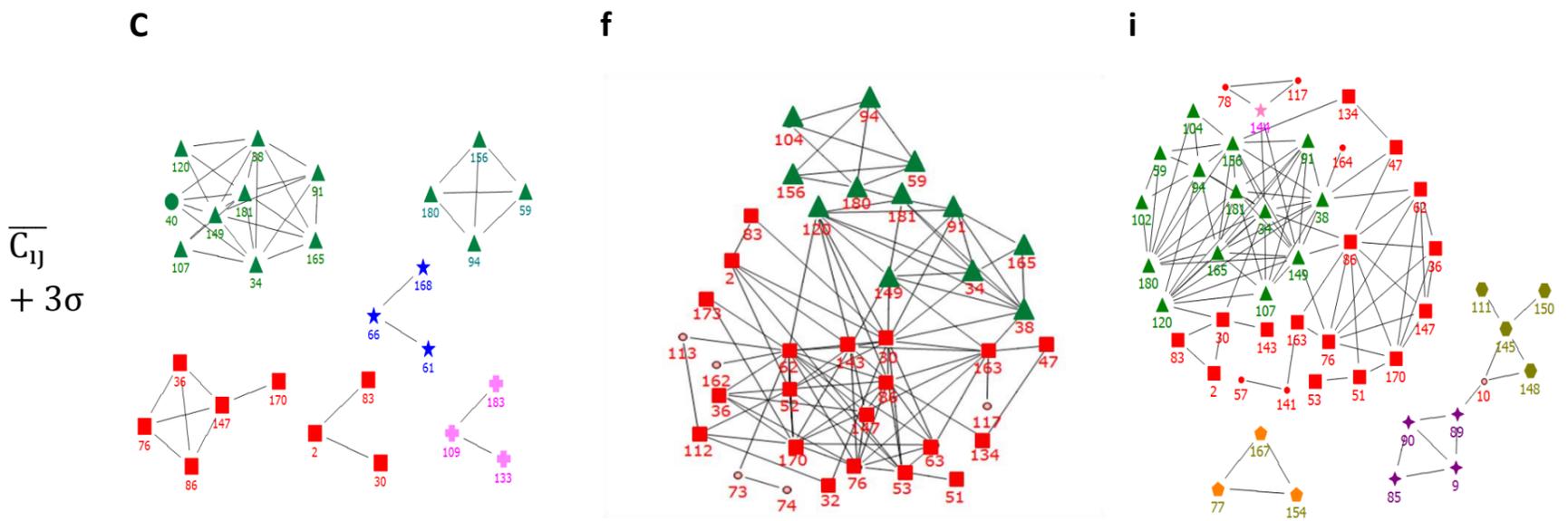

Figure 2 | The threshold networks with three different thresholds for three different periods: before the crisis a, θ=0.3313, b, θ=0.4304, c, θ=0.5295, during the crisis; d, θ=0.4665, e, θ=0.5805, f, θ=0.6945, after the crisis; g, θ= 0.2508, h, θ= 0.3535, i, θ= 0.4562. We visualize only the largest cluster because the figure would become crowded if we were to plot all the vertices of the figure. At the high threshold, the size of the largest cluster is the biggest during the crisis. At the threshold $\theta = \overline{C_{IJ}} + 3\sigma$, we observed the obvious clustering of the companies to the sectors. Before the crisis, there are clusters which belong to finance (green triangle), heavy industry and construction (red Square), iron and metal products (magenta plus), and non-metallic mineral products (blue star). However, during the crisis, there is one cluster that belongs to the financial sector and one to the heavy industrial and shipbuilding sectors (construction). After the crisis, the largest one is the financial sector and the heavy industrial and shipbuilding sectors (construction), and the others consist of the electronics and automobile and transport equipment (diamond orange), respectively. A star symbol indicates Samsung Electronics (magenta, 144).

The mean degree of the largest cluster is a decreasing function for the value of the threshold (Fig. 1S). This tends to decrease as the threshold increases since the number of connected vertices decreases with the increasing threshold. At a fixed value of the threshold, the mean degree during the crisis is higher than during other periods, which indicate more interaction among the indices in this period. In the threshold network, all vertices do not have the same number of edges. The degree distribution P(k) characterizes the number of edges of a vertex, which gives the probability that a selected vertex has exactly k edges. In real networks[16-18] such as the World Wide Web, the Internet, cell networks, the degree distribution follows a power law, $P(k) \sim k^{-\gamma_d}$. We observe in the largest cluster of the threshold network that the degree distribution for a certain interval of the threshold follows the power law (Fig. 2S). The characterization of the degree distribution for different values of the threshold is given in Table S2. We observed that as θ increases from 0.4125 to 0.5 for before and after the crisis, the degree distribution follows power law while for during the crisis the interval of θ is [0.575, 0.65]. For other correlation thresholds, the distributions of the degrees do not have any scale-free structure. The high value of the threshold for the power law degree distribution during the crisis in comparison to other periods is due to the high edge density up to the high threshold. Since the indices of the companies were strongly correlated during the crisis, the vertices carry a high-edge density.

The average clustering coefficient decreases as the threshold increases, shown in Figure 3S by the logarithmic scale. We observed that within a certain range of the threshold θ ∈ [0.25,0.5] before and after the crisis and during the crisis $\theta \in [0.4, 0.675]$, the average clustering coefficient shows the scaling law, $C(\theta) \sim \theta^{-\alpha}$, where the exponent α are 3.0(6) ($r^2 = 0.88$) before the crisis, 3.5(5) ($r^2 = 0.97$) during the crisis, and 2.1(2) ($r^2 = 0.92$) after the crisis. Furthermore, we observed that the average clustering coefficient, $C(\theta)$, decreases rapidly during the crisis. The clustering coefficients depend on the edges and the neighbors of the vertex[8]. Since the threshold network is highly clustered during the crisis, the edge density is higher and, consequently, the decreasing rate of the edge density with the thresholds is faster than in other periods. As a result, the average clustering coefficients decay sharply during the crisis. On the other hand, the networks are loosely connected after the crisis. For this reason, the decreasing rate of the edge density is slower and, hence, the exponent α is smaller than other periods.

**a**

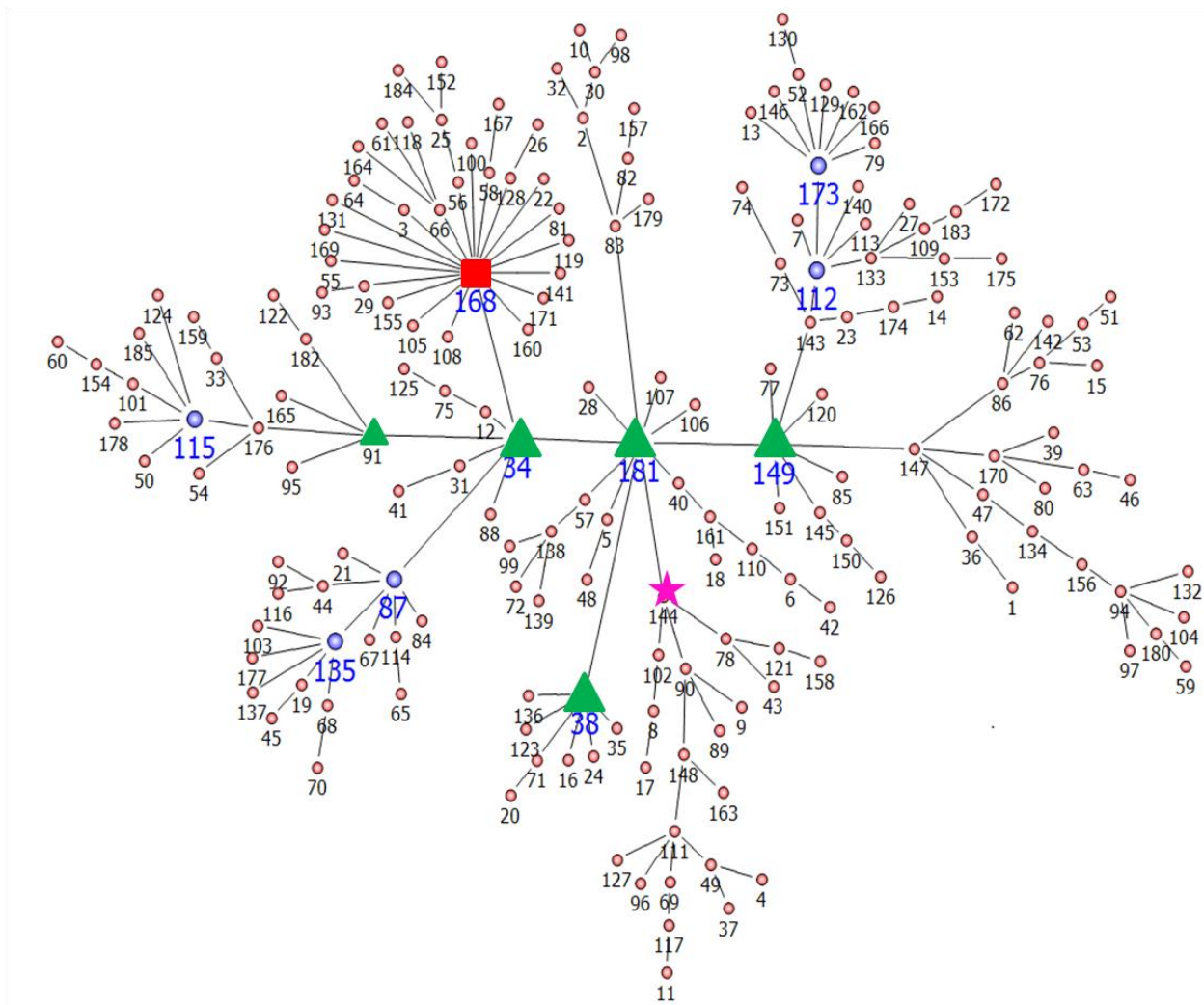

**b**

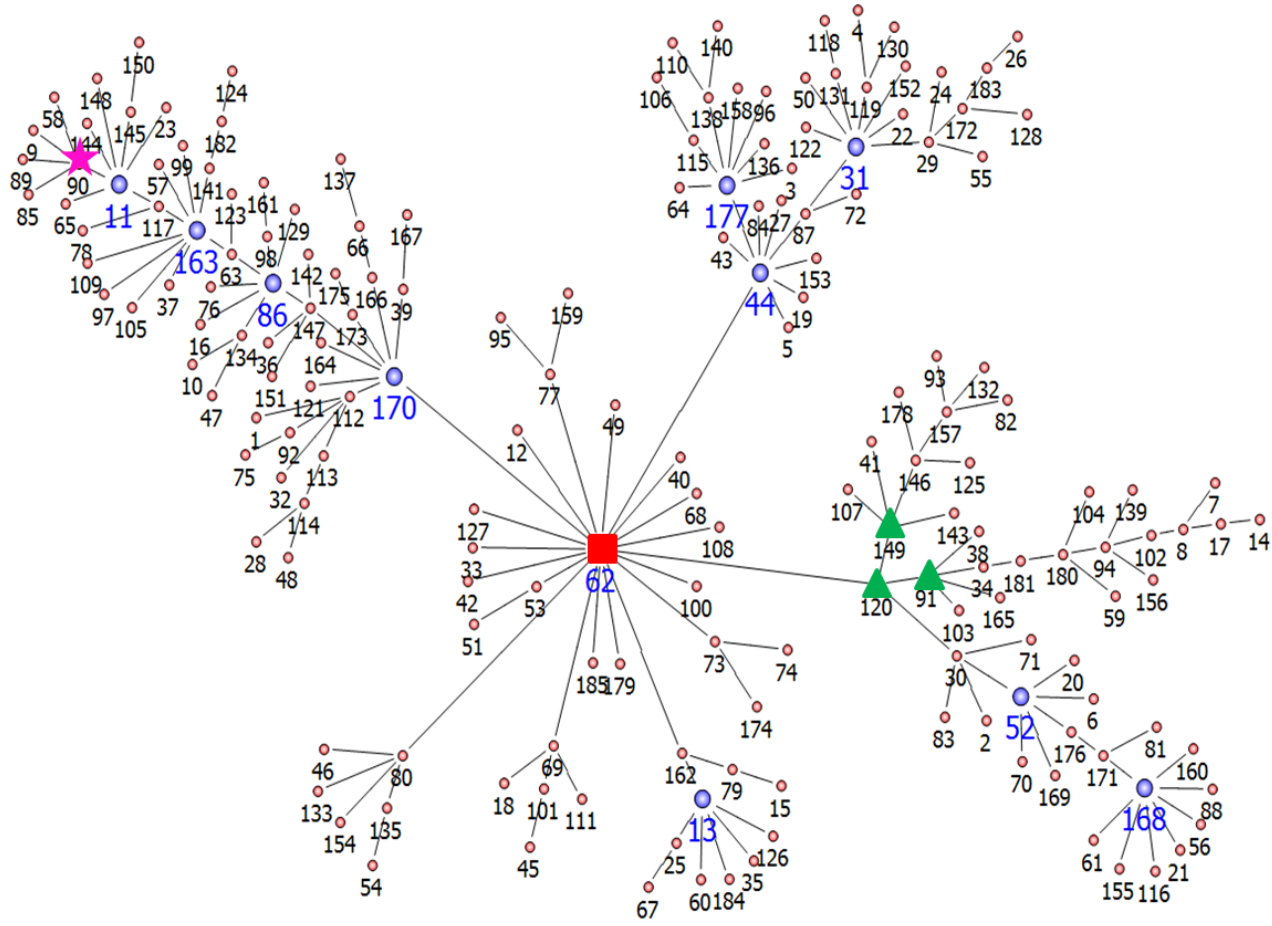

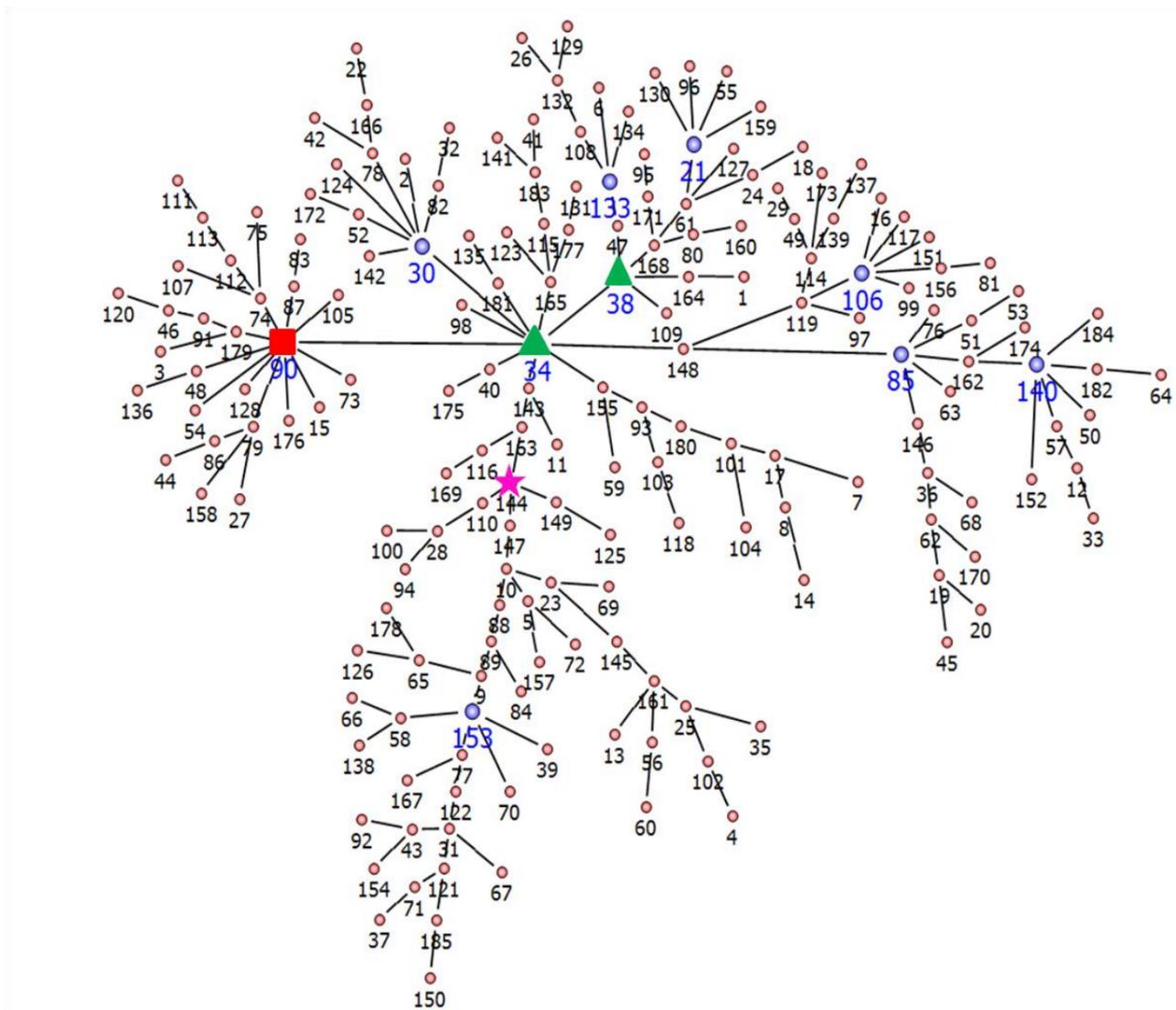

Figure 3 | The minimum spanning trees obtained at three periods; a, before the crisis; b, during the crisis; c, after the crisis. The companies are marked by the small circle and numbers, and the edges represent the

distances between the companies. The square symbol represents the hub vertex and the triangles correspond to the companies that belong to the financial sector. The star is Samsung Electronics (144) which is the most influential company in the Korean stock market. We use the numbers on the vertex because if we label the company name with text, the figure will be crowded. The names of the companies corresponding to the numbers are provided in a supplementary Excel file. During the crisis, Hanjin Heavy Industries and Construction (HHIC), which corresponds to number 62 comes out as a center hub. There are several local hubs before and after the crisis. The HHIC, a legend in the heavy industries, is one of the world's best heavy industry specialists equipped with digital technology. During the crisis, this may play an important role in resisting the crisis shock. As a result, it becomes a central hub on the MST during the crisis. The big numbers on the MST correspond to the top ten vertices having the high degrees.

We construct asset trees in order to visualize a complex network before, during, and after the global financial crisis (Fig. 3). The big numbers on the MST correspond to the top ten vertices having the high degrees (Fig. 3). The top ten hubs in each MST list is in table 3S. Before the crisis, the top hub is the Ssangyong Cement Industry (number 168). The four financial companies (the vertex numbers are 34, 38, 149, and 181) are located around the center of the MST (Fig 3a). We observe the topological transitions of the asset tree during the crisis where Hanjin Heavy Industries and Construction (20[th] largest company in Korea, the vertex number 62) comes out as a center hub like a dragon king (Fig. 3b). This is the dynamic topological transition from the asset tree having several local hubs before the crisis to a super hub-like asset tree during the crisis. The companies in the financial sector are forming only a minor branch in the MST. After the crisis, this super-hub disappeared, as shown in Figure 3c. These types of transitions were also found in the German stock exchange (FSE)[20]. The superstar-like hub indicates the unstable market state, and hence a large group and compact structure of the indices to resist a crisis shock. After the crisis, two companies (number 34 and 38) in the financial sector become the center of the star-like MST. In this period, the main hub is Hyundai Motor (number 90). Throughout every period, the Samsung Electronics (the Korean top company) is located in the minor branch in the MST. On the other hand, before and after the crisis, the tree containing several hubs indicates the meta-stable market state and by extension, several groups and large structure of indices. The crisis is like a natural disaster. At the time of the disaster, group formations to protect the disaster shock indicate the unstable state.

During the last few years, much attention has been devoted to the degree distribution of graphs called scale-free trees where the degree distribution obeys power law. Scale-free networks have extensively studied in economy and finance[23-27]. The degree distribution, $P_{MST}(k)$, in the MST follows power law, $P_{MST}(k) \sim k^{-\gamma_m}$ where k is the degree of the vertex. The exponent, $\gamma_m$, implies that the mean size or second moment of the degree would diverge in the infinite degree limit. The MST shown in Figure 3a follows scaling law with the exponent $\gamma_m = 1.98(36)$ ($r^2 = 0.96$) which is lower than in normal periods (Fig. 3c). However, at the crisis period, we have $\gamma_m = 1.84(45)$ ($r^2 = 0.92$), which is smaller than those as usual period. We observe similar behaviors of the degree exponents for the threshold network. The lower value of the exponent during the crisis is due to the shrinking of the tree, which means that the nodes of the high degree are increased. Further, we calculate the average tree length $L$ in the MST (see Method Summary). The measured average tree lengths are $L = 1.05, 0.92$, and $1.1$ for before, during, and after the crisis, respectively. During the crisis, the size of the MST is smaller than the other period which indicates that the market behaviors tightly correlated manners. Moreover, the lower value of the exponent before the crisis than as usual periods is a precursor of the market crash. After the crisis, the exponent $\gamma_m = 2.2(9)$ ($r^2 = 0.88$) which is close to the usual value for Korean stock market, indicating a more uniform degree. So, we can divide the Korean market into three states: metastable (before crisis), unstable (during crisis), and completely stable (after crisis). The exponent $\gamma_m$ varies from 2.1 to 2.2 as the usual time that we found by analyzing data from 2002 to 2005. The slight deviation of exponent $\gamma_m$, representing the normal period for before the crisis from normal period, is a red alert of the market crash. The exponent that we found for the MST from the Korean stock market for the normal and crash periods are analogous to the exponents of New York stock exchange (NYSE) in the financial crass of 1986[23]. We can then suppose that the Korean stock market is complementary with the New York stock exchange.

A dendrogram gives an alternative representation of the network that shows the full hierarchical structure[8,9]. At the first level of the dendrogram, there are N-singleton clusters. As one moves the vertical scale of the dendrogram, the clusters are combined until all vertices are contained in a single community at the top of the dendrogram. The hierarchical structure in the financial markets has been analyzed for the Dow Jones Industrial Average (DJIA) and the Standard and Poor's 500[10]. The tree-like dendrogram provides a meaningful economic taxonomy. We apply the average linkage hierarchical clustering algorithm to the distance matrix to produce the dendrogram. We construct a tree-like dendrogram in the Korean stock market during, before, and after the crisis. Here, we take 30 stocks because if we use all of the stocks, the figure will be crowded and it will be difficult to explain the figure clearly (Fig. 5S). Here, our focus is on the height of the nearest clusters in hierarchical network in all periods. If we restrict the range of the distance, $d_{ij} \leq 1$, we observe that the number of the pairs of vertices in this range is 31(before crisis), 82(during crisis), and 27(after crisis) out of 184 pairs(not shown in Fig 5S). The number of pairs between $1.0 \leq d_{ij} \leq 1.2$ is 102 (before crisis), 80(during crisis) and 66 and for the distance, $d_{ij} > 1.2$, the numbers are 51(before crisis), 22(during crisis), and 91(after crisis). We observe that the height between the nearest clusters decreases in the hierarchical network during the crisis. Of course, these decreasing trends are not true for all pairs of the companies. Throughout all the examined periods, we found that some companies were forming tight bonds with each other. The cluster before the crisis also reorganizes with the new company during and after the crisis (Fig. 5S). We propose then that when a crisis shocks the stock market, companies can protect themselves from this by making a loose or strong bond to other companies or finding some other factors to relax the shock. The cophenetic correlation for a hierarchical network is measured by the linear correlation coefficient between the cophenetic distances induced from the tree and the distances used to construct the tree. The cophenetic distance between two observations is represented in a dendrogram by the height of the link when those two observations are joined for the first time. The cophenetic correlation coefficient (CCC) measures the correlation between the dissimilarity network and the hierarchical network[29]. We obtained CCC values in the three periods as 0.8165 before the crisis, 0.8959 during the crisis and 0.788 after the crisis. The CCC value during the crisis is comparatively higher than before and after crisis which indicates that the hierarchy increases during the crisis[8,11].

The network analysis of the correlation matrices provides information about the formation of clusters of the indices. The network topologies indicate that the indices of the companies highly correlated each other during the crisis. We found the scaling behaviors in the degree distribution and the clustering coefficient in the threshold network. The size of the network shrinks in the MST and HN during the crisis. The network structures change heavily during the crisis. Further understanding of the network structures in the stock market during the crisis will help in predicting the influence of these impacts on the market.

Methods Summary

**Data sets**. We analyze the time series of the daily closing stock prices (local indices) of 185 Korean companies belonging to the KOSPI 200 in the period from June 2, 2006 to December 30, 2010. These 185 companies survived and kept their company names in this period. We divide the time into three periods based on volatility. We consider the period from June 2, 2006 to November 30, 2007 as before the crisis consisting 388 days, from December 3, 2007 to June 30, 2009 as during the crisis consists 412 days and from July 1, 2009 to December, 30 2010, which consists of 389 as after the crisis. We take the period during the crisis as the period with a highly fluctuating volatility. During the crisis, the Lehman Brothers went bankrupt in November 15, 2008. This collapse spurred the global financial crisis. The financial crisis of 2007-2009 is known as the worst financial crisis since the Great Depression in the 1930s, and originated in the United States before spreading all over the world[21].

**Cross-correlation coefficients.** Let the closing stock price of a company be $I_i(t)$ at a time *t*. For each window, we calculate the normalized returns of the stock price of each company as $r_i(t) = [ln\, I_i(t) - ln\, I_i(t-1)]/\sigma_i$, where $\sigma_i$ is the standard deviation of the time series. Then, we calculate the cross-correlation coefficients between the return time series as $C_{ij}(t) = <r_i(t)r_j(t)> - <r_i(t)><r_j(t)>$.

**Mean degree.** The degree of the vertex *i* is $k_i = \sum_{j \neq i} e_{ij}$, which is the total number of connections in the network[22], where $e_{ij} = 1$ if the vertex *i* and *j* are connected, $e_{ij} = 0$, otherwise. The mean degree is the average of $k_i$ over all vertices in the network.

**Average clustering coefficient.** The clustering coefficient of a vertex *i* is defined as $C_i = 2m_i/n_i(n_i - 1)$, where $n_i$ denotes the number of neighbors of the vertex *i*, and $m_i$ are the edges existing between the neighbors of vertex *i*. $C_i$ is equivalent to 0 if $n_i \leq 1$. The average clustering coefficient at a specific threshold for the entire network is defined as the average of $C_i$ over all the nodes in the network, that is, $C = \frac{1}{N}\sum_{i=1}^{N} C_i$ where *N* is the total number of the vertex.

**Minimal spanning tree.** A minimum spanning tree is constructed by calculating the distance matrix of the indices[23-25]. The distance matrix is $d_{ij} = \sqrt{2(1 - C_{ij})}$, where $d_{ij}$=0 if the price time series of the index *i* and *j* are perfectly correlated, and $d_{ij}$=2 if the price time series of the index *i* and *j* are perfectly anti-correlated. The MST has been built following Kruskal's algorithm in order to find the *N-1* most important correlated pairs of the indices among the *N(N-1)/2* possible pairs.

**Average tree length.** The size of the tree characterizes by the average tree length in the MST. We defined the average tree length as[16-18]

$$L(t) = \frac{1}{N} \sum_{<i,j>} d_{ij}^{MST}$$

where *N* is the total number of the vertex in the tree and $d_{ij}^{MST}$ is the shortest path length between two vertex *i* and *j*.

**Cophenetic correlation coefficient.** The Cophenetic correlation coefficient CCC is defined as[11,29]

$$CCC = \frac{[\sum_{i<j}(d_{ij}-\bar{d})\times(c_{ij}-\bar{c})]}{\sqrt{[\sum_{i<j}(d_{ij}-\bar{d})^2][(c_{ij}-\bar{c})^2]]}},$$

where $d_{ij}$ and $\bar{d}$ are the element and the average elements of the distance matrix, and $c_{ij}$ and $\bar{c}$ are the elements and average elements of the cophenetic matrix, respectively.

Supplementary Information

Supplementary Tables

**Table S1 |** Statistical properties for the distribution of the cross-correlation coefficient obtained for before during, and after the crisis.

| Period | Mean Cross-Correlation | Standard deviation | Skewness | Kurtosis | Mean volatility |
|---|---|---|---|---|---|
| Before | 0.23 | 0.098 | 2.80 | 27.22 | 0.019 |
| During | 0.35 | 0.11 | 1.04 | 13.98 | 0.025 |
| After | 0.15 | 0.10 | 3.80 | 35.37 | 0.017 |

**Table S2 |** Liner least squares estimate the degree exponent $\gamma_d$ for different values of the correlation threshold. The lower $\gamma_d$ indicates a high-edge density for many vertices while the higher $\gamma_d$ indicates a high edge density of a few vertices. The lower exponent for the lower threshold implies that most of the vertices have uniform edge density. We observe that the degree exponent $\gamma_d$ increases with θ after the crisis, but exhibits dissimilar behaviors in the periods before and during the crisis. This may be the meta-stable and unstable condition of the market during the crisis, respectively. The smaller values of the degree exponent $\gamma_d$ during the crisis are due to high edge density of many vertices and consequently large cliques.

| Before the crisis | | During the crisis | | After the crisis | |
|---|---|---|---|---|---|
| θ | $\gamma_d$ | θ | $\gamma_d$ | θ | $\gamma_d$ |
| 0.4125 | 0.85(34) | 0.575 | 0.75(48) | 0.4125 | 0.97(39) |
| 0.425 | 1.33(44) | 0.60 | 0.86(66) | 0.425 | 1.01(51) |
| 0.45 | 0.94(44) | 0.6125 | 0.74(58) | 0.45 | 1.23(51) |

**Table S3 |** The top ten hubs in the minimum spanning trees in three periods. Before the crisis, the several companies in the financial sector belong to the top ten hubs. However, during the crisis, the companies in heavy industries occupy the top ten hubs. After the crisis, there are no dominant sectors occupying the top ten hubs.

| Ranking | Before the crisis | | During the crisis | | After the crisis | |
|---|---|---|---|---|---|---|
| | Vertex number -Company name | Degree | Vertex number -Company name | Degree | Vertex number -Company name | Degree |
| 1 | 168-Ssangyong cement industry | 20 | 62-Hanjin heavy industries and construction | 20 | 90-Hyundai Motor | 12 |
| 2 | 181-Woori investment securities | 11 | 44-Dongbu steel Co. Ltd. | 9 | 34-Daewoo securities | 10 |
| 3 | 173-Taihan electric wire | 8 | 163-SK holdings | 9 | 30-Daelim industrial Co. Ltd. | 7 |
| 4 | 149-Samsung securities | 8 | 31-Daesang | 8 | 85-Hyundai glovis Co. Ltd. | 6 |
| 5 | 34-Daewoo securities | 8 | 168-Ssangyoung cement industries | 8 | 140-S&T holding Co. Ltd. | 6 |
| 6 | 38-Daishin securities | 7 | 170-STX offshore and shipbuilding | 8 | 21-Binggrae Co. Ltd. | 5 |
| 7 | 87-Hyundai hisco Co. Ltd. | 7 | 177-Unio steel Co. Ltd. | 8 | 38-Daishin securities | 5 |
| 8 | 135-Posco coated colour steel | 7 | 11-LG electronics | 7 | 106-Korea industrial Co. Ltd. | 5 |
| 9 | 112-Kumho industrial Co. Ltd. | 6 | 86-Hyundai heavy industries Co. Ltd. | 6 | 153-Seah bestell corp. | 5 |
| 10 | 115-Kwang dong pharmaceutical Co. Ltd. | 6 | 52-Doosan construction and Engineering | 6 | 133-Poongsan holding corp. | 4 |

Supplementary Figures

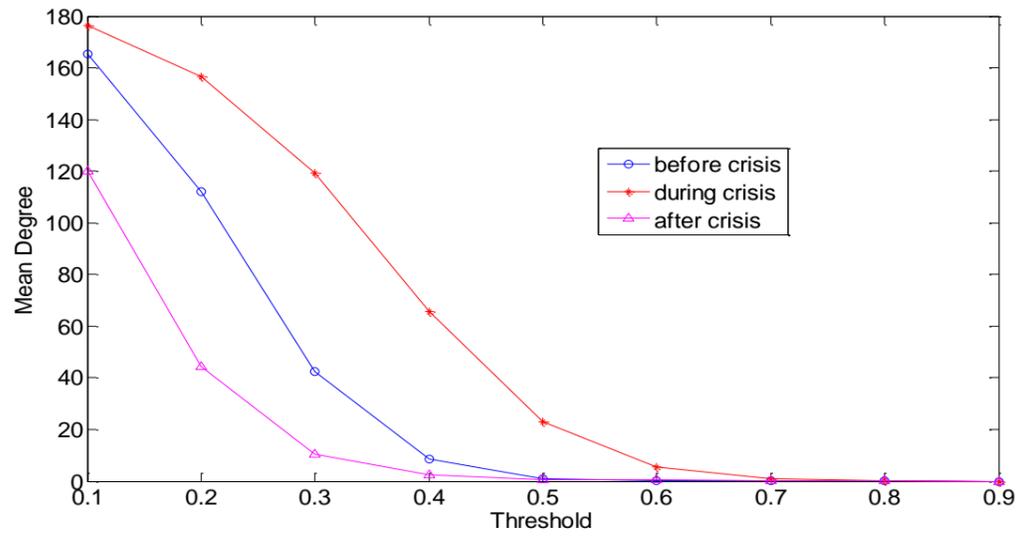

**Figure S1 |** The mean degree for before, during, and after the crisis at the different thresholds. The mean degree is decreasing function against the threshold. The high mean degree is observed during the crisis due to a strong interaction among the indices (the unstable market state). The less mean degree after the crisis indicates less interaction among the indices and a more stable market state.

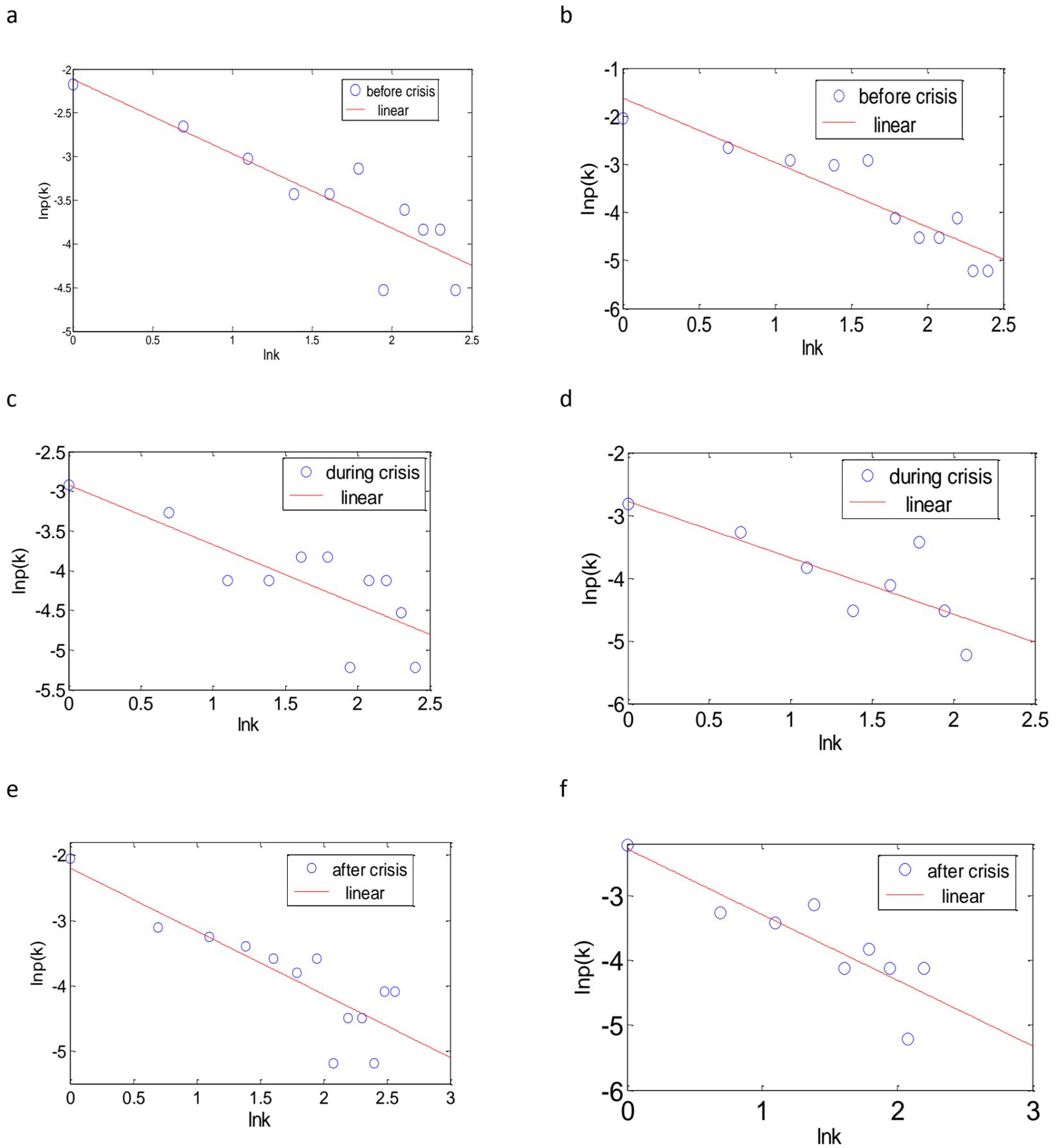

**Figure 2S |** The log-log plots of the degree distribution in the threshold network are represented as a function of the degree for before the crisis, a, θ=0.4125, b, θ=0.425, during the crisis, c, θ=0.575, d, θ=0.6, after the crisis, e, θ=0.4125, f, θ=0.425. The degree exponents obtained by the least square fits are given in the table S2.

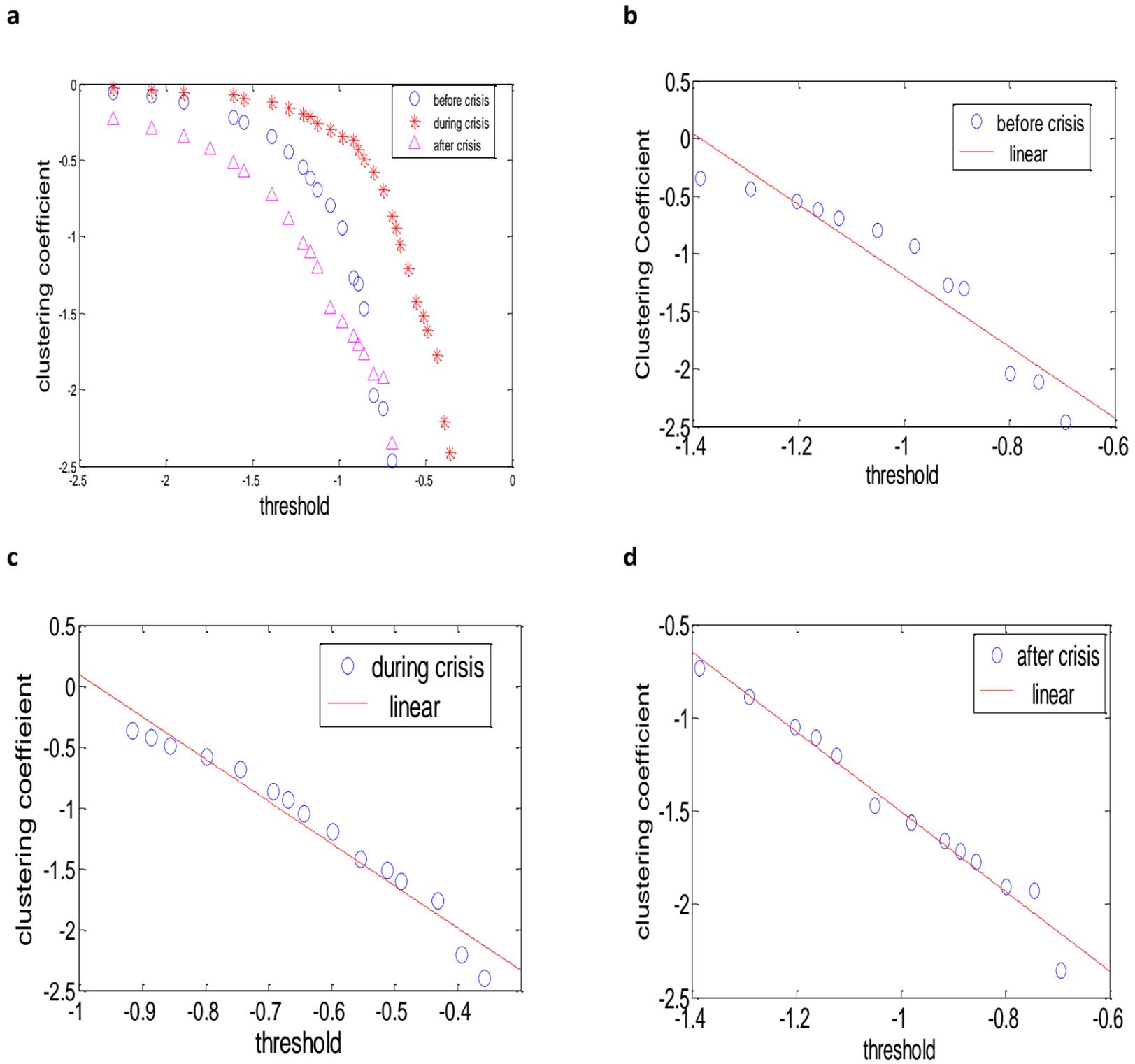

**Figure S3** | a, The average clustering coefficients obtained as a function of the threshold in the threshold networks. Log-log plots of the clustering coefficients represent a function of the threshold θ, b, before the crisis, c, during the crisis, d, after the crisis, respectively.

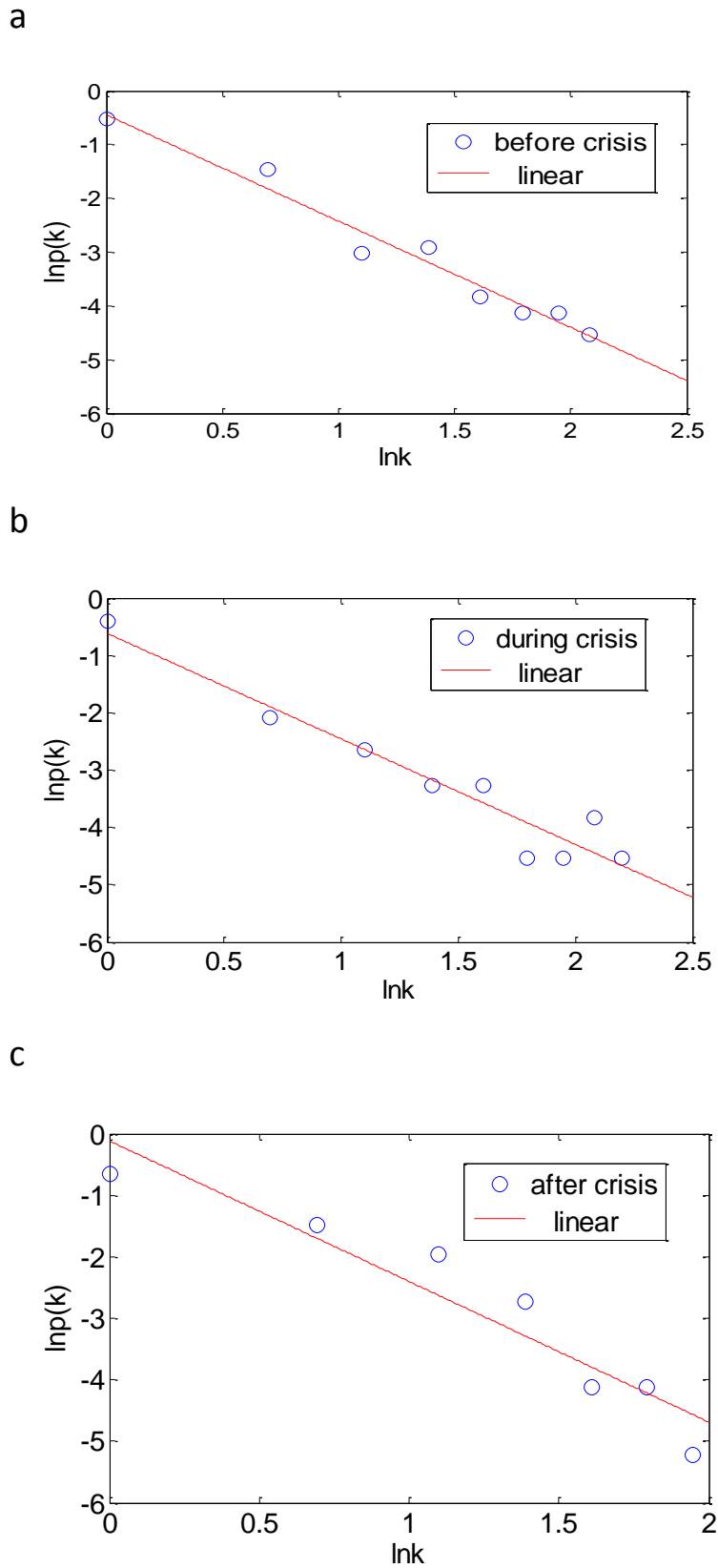

**Figure 4S** | The log-log plots of the degree distribution function represent a function of the degree in the MST for a, before the crisis, b, during the crisis, c, after the crisis, respectively. We obtained degree exponents $\gamma_m \approx 1.98(36)$ for before the crisis, $\gamma_m \approx 1.84(45)$ for during the crisis, and $\gamma_m \approx 2.2(9)$ for after the crisis.

a

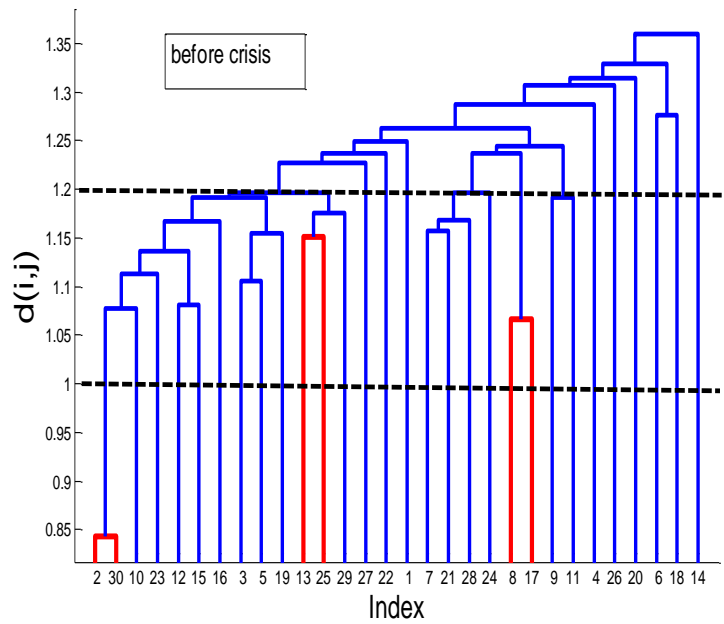

b

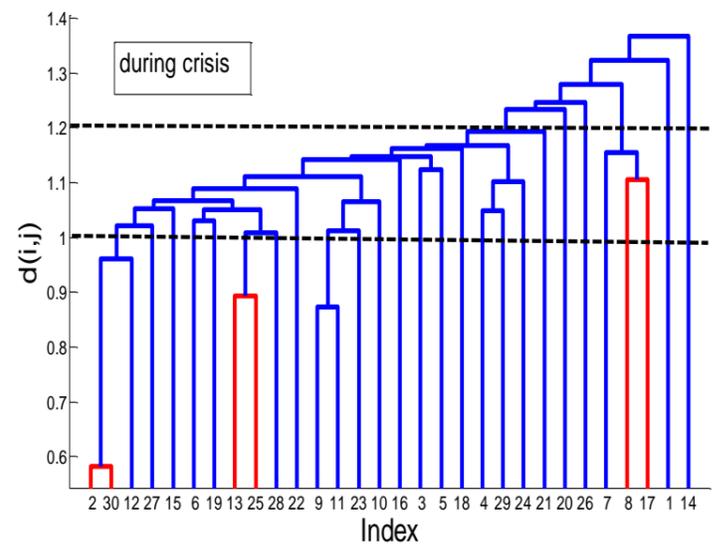

c

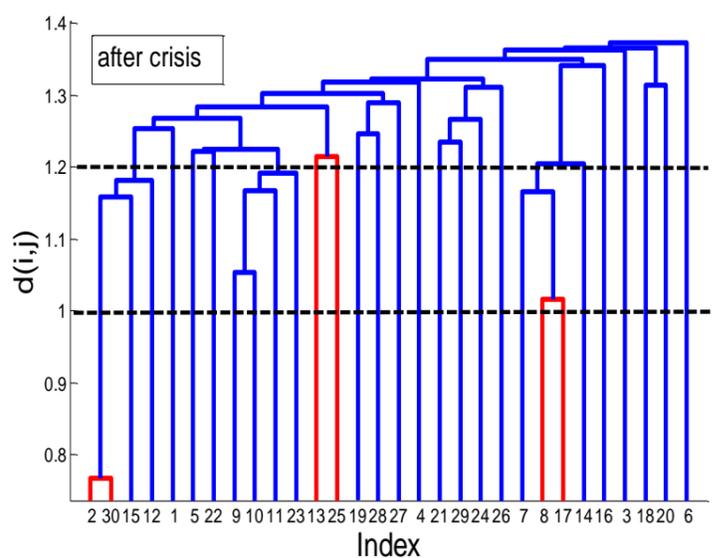

**Figure 5S |** Dendrogram representations of financial indices, a, before the crisis, b, during the crisis, and c, after the crisis. The numbers representing the individual company are given in a supplementary Excel file. The height of the dendrogram decreases during the crisis while increases before and after the crisis. The companies at the sector of construction (2,30), communication (8,17), and medical supply (13,25) make a tight bond exist during every period.